\documentclass[conference]{IEEEtran}
\IEEEoverridecommandlockouts

\usepackage{cite}
\usepackage{amsmath,amssymb,amsfonts}
\usepackage{algorithmic}
\usepackage{graphicx}
\usepackage{textcomp}
\usepackage{csquotes}
\usepackage{hyperref}
\usepackage{xcolor}
\def\BibTeX{{\rm B\kern-.05em{\sc i\kern-.025em b}\kern-.08em
    T\kern-.1667em\lower.7ex\hbox{E}\kern-.125emX}}

\usepackage{hyperref}
\hypersetup{
colorlinks=true,
linkcolor=black
}

\usepackage{caption} 
\usepackage{tabularx}
\usepackage{float}
\usepackage{array}  
\usepackage{booktabs} 
\usepackage{multirow}
\usepackage{makecell}

\usepackage{color}

\usepackage{bbding}

\usepackage{amsfonts,amssymb} 

\usepackage{subfigure} 

\begin{document}

\title{Sound-Based Recognition of Touch Gestures and Emotions for Enhanced Human-Robot Interaction
}

\author{%
\IEEEauthorblockN{%
Yuanbo Hou\IEEEauthorrefmark{1},
Qiaoqiao Ren\IEEEauthorrefmark{2},
Wenwu Wang\IEEEauthorrefmark{3}, 
Dick Botteldooren\IEEEauthorrefmark{1}
}\\ 
  
\IEEEauthorblockA{%
\IEEEauthorrefmark{1}WAVES, Ghent University, Belgium
\IEEEauthorrefmark{2}AIRO, Ghent University-IMEC, Belgium
\IEEEauthorrefmark{3}CVSSP, University of Surrey, UK
}
}

\maketitle 

\begin{abstract}

Emotion recognition and touch gesture decoding are crucial for advancing human-robot interaction (HRI), especially in social environments where emotional cues and tactile perception play important roles. However, many humanoid robots, such as Pepper, Nao, and Furhat, lack full-body tactile skin, limiting their ability to engage in touch-based emotional and gesture interactions. In addition, vision-based emotion recognition methods usually face strict GDPR compliance challenges due to the need to collect personal facial data. To address these limitations and avoid privacy issues, this paper studies the potential of using the sounds produced by touching during HRI to recognise tactile gestures and classify emotions along the arousal and valence dimensions. Using a dataset of tactile gestures and emotional interactions from 28 participants with the humanoid robot Pepper, we design an audio-only lightweight touch gesture and emotion recognition model with only 0.24\textit{M} parameters, 0.94\textit{MB} model size, and 0.7\textit{G} FLOPs. Experimental results show that the proposed sound-based touch gesture and emotion recognition model effectively recognises the arousal and valence states of different emotions, as well as various tactile gestures, when the input audio length varies. 
The proposed model is low-latency and achieves similar results as well-known pretrained audio neural networks (PANNs), but with much smaller FLOPs, parameters, and model size.

\end{abstract}


\vspace{-0.2cm}
\section{Introduction}
\label{sec:intro}

\vspace{-0.05cm}
Robots' perception of tactile gestures and emotions is integral to the development of advanced human-robot interaction (HRI), especially in social environments, where understanding and responding to emotional cues is critical for meaningful interactions \cite{su2023recent, spezialetti2020emotion, marques2022affect}. Moreover, real-time emotion states and touch gesture recognition can significantly improve the naturalness and effectiveness of HRI \cite{villani2020humans}\cite{6926297}.

Previous work \cite{basdogan2020review} has attempted to decode touch gestures from tactile signals, but a major limitation is that many robots lack skin that can sense touch throughout the body. Most research on humanoid robotics values touch less than vision or audio signals. Robots like Pepper \cite{pandey2018pepper}, NAO \cite{NAO}, and Furhat \cite{al2012furhat} do not have extensive tactile sensors covering the entire body \cite{dahiya2009tactile}. This limitation hinders the development of HRI systems that decode gestures and emotions based on tactile perception. To this end, tactile sensors embedded in the surface of robots are used to decode touch gestures based on pressure patterns, thus providing a direct tactile-based interaction method \cite{yuan2017gelsight}. However, tactile sensors \cite{meribout2024tactile} can be intrusive and are often limited by the need for different body contacts.

In addition to tactile-based gesture recognition, most gesture decoding studies adopt vision-based methods, which use depth cameras \cite{ren2011depth} or optical sensors \cite{xia2019vision} to capture human gestures and emotions. These vision-based models perform well in controlled environments. However, they often face significant challenges in dynamic real-world environments \cite{shotton2011real}, e.g., occlusion of certain parts of the face/body or changes in lighting conditions can seriously affect the effectiveness of the vision-based models \cite{noroozi2018survey}. Moreover, the vision-based models usually require relatively high computational resources, and collection of training data. However, it is difficult to avoid issues related to privacy and general data protection regulation (GDPR) during data collection. These limitations restrict the scalability and practicality of vision-based methods.

To recognize gestures and emotions, researchers have also explored auditory signals and demonstrated the feasibility of sensing touch gestures on the surface of mobile devices based on acoustic signals \cite{sun2018vskin}. Auditory stimuli are viewed to be potent triggers of affective responses, and specific sounds, e.g., tapping sounds, can affect people's emotional states and behaviours \cite{tajadura2015sonification}. Moreover, speech-based emotion recognitions have been successfully applied to adjust the behaviour of social and educational robots to adapt to current social emotions. Furthermore, the recognition of emotions such as happiness, sadness, and anger has been widely carried out using features such as the rhythm, pitch, and intensity of speech \cite{cowie2001emotion, eyben2009openear, gasteiger2024scoping}.

Emotions are typically analysed along two independent dimensions (arousal and valence), like in Russell's model \cite{russell1989cross}\cite{russell1980circumplex}, and neuroimaging studies \cite{citron2014emotional}\cite{colibazzi2010neural} support these two-dimensional representations. Hence, the arousal-valence dimensional model (AVDM) is more commonly used for sound-related affective computing, e.g., soundscape studies \cite{botteldooren2011understanding}\cite{botteldooren2013soundscape}, than the discrete emotion model for single discrete entities \cite{annoyance}\cite{soundscaper}. Hence, this paper uses AVDM to analyse touch-related emotions.

This paper makes the first attempt to recognise touch gestures and emotions based on sounds produced during HRI. Although previous studies have shown the feasibility of sensing touch gestures based on acoustic signals, they did not further study the decoding of different gestures from participants. Whereas most of the previous auditory-based human emotion recognition related to robots is based on speech signals, this paper utilises the non-speech sounds produced during HRI.

\begin{figure}[t]
    \centering
    \setlength{\abovecaptionskip}{0.15cm}    
	\setlength{\belowcaptionskip}{-0.5cm} 
    \includegraphics [width=0.85\columnwidth]{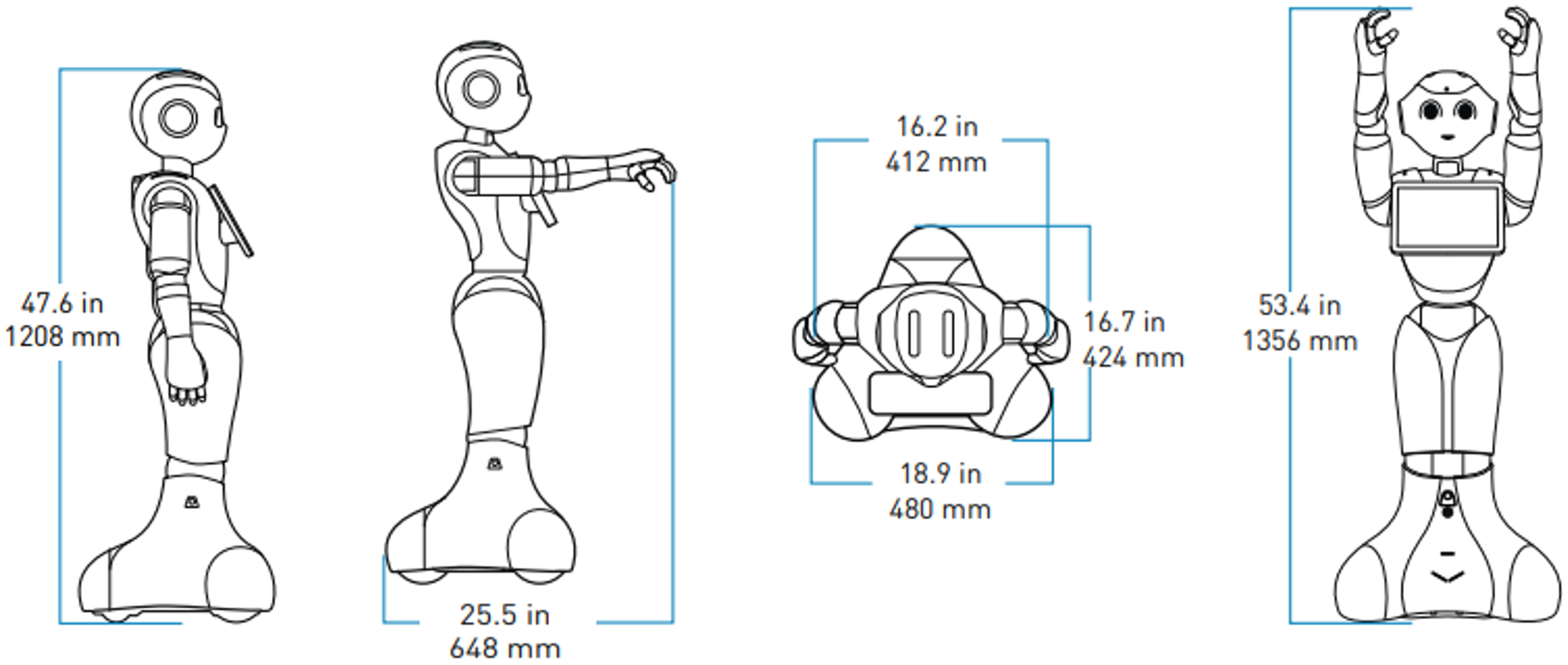}
    \caption{The robot Pepper's physical information$^1$.}
    \label{pepper}
\end{figure}

\section{Background}\label{background}

\begin{figure}[H]
    \centering
        \includegraphics [width=0.7\columnwidth]{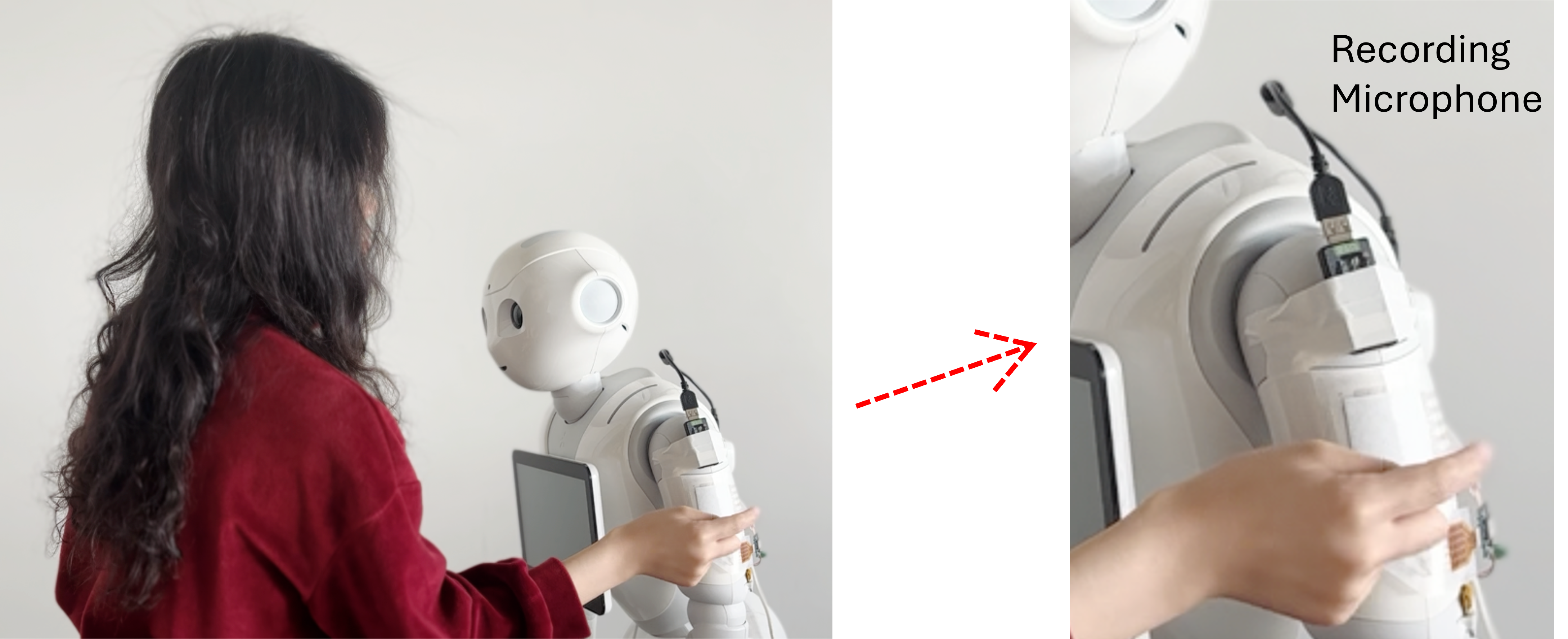}
    \caption{The participant interacts with the robot Pepper.}
    \label{interaction}
\end{figure}

Pepper is an interactive robot developed by SoftBank Robotics \cite{pandey2018pepper}, as shown in Fig.~\ref{pepper}; more technical specifications, please see here\protect\footnote{\url{https://support.aldebaran.com/support/solutions/articles/80000958735-pepper-technical-specifications}}.
In the data collection experiment, participants are asked to express 6 gestures (\textit{hold}, \textit{pat}, \textit{poke}, \textit{tickle}, \textit{tap}, \textit{rub}) and 10 emotions by touching Pepper's left forearm with spontaneous movements. 
The sounds produced by these touch movements are recorded by a microphone (the black device in Fig. \ref{interaction}) to form the sound dataset in this paper.

According to the Circumplex Model \cite{russell1980circumplex}, the distribution of the 10 emotions involved in arousal-valence dimensions is shown in Fig. \ref{russel_model}. 
\textit{Happiness} and \textit{surprise} occupy the high arousal, positive valence quadrant (Q1);
\textit{anger}, \textit{fear} and \textit{disgust} are located in the high arousal, negative valence quadrant (Q2); \textit{sadness} and \textit{confusion} are located in the low arousal, negative valence quadrant (Q3); \textit{comfort} and \textit{calming} belong to the low arousal, positive valence quadrant (Q4); the neutral emotion, \textit{attention}, is located at the origin (Q0).

\vspace{-0.2cm}

\section{Proposed Method}

Our ultimate goal is to develop a sound-based model that can be embedded into a robot to perceive touch gestures and emotions. To this end, this section first analyzes the limited availability of computing resources in the robot brain system and then designs a lightweight model tailored for the robot. 


\subsection{Robot's brain system computing resources}\label{guidelines}

Pepper's Brain system consists of an Intel ATOM® E3845 processor and 4 GB of DDR3 RAM. The documentation\protect\footnote{\url{https://www.intel.com/content/dam/support/us/en/documents/processors/APP-for-Intel-Atom-Processors.pdf}} provided by Intel shows that E3845 processor's Floating Point of Operations (FLOPs) is 11.46$G$ per second.
To deploy a model in Pepper's brain system, the model's FLOPs should not be greater than 11.46$G$; otherwise, it cannot be run on Pepper. Considering this restriction, the model to be deployed should meet the following requirements: 1) lightweight with few parameters; 2) small FLOPs and low processing latency; 3) able to process varying-length audio clips. Next, we will use these three points as guidelines to design the model.

\begin{figure}[t]
    \centering
     \setlength{\abovecaptionskip}{0.15cm}    
	\setlength{\belowcaptionskip}{0cm} 
        \includegraphics [width=1\columnwidth]{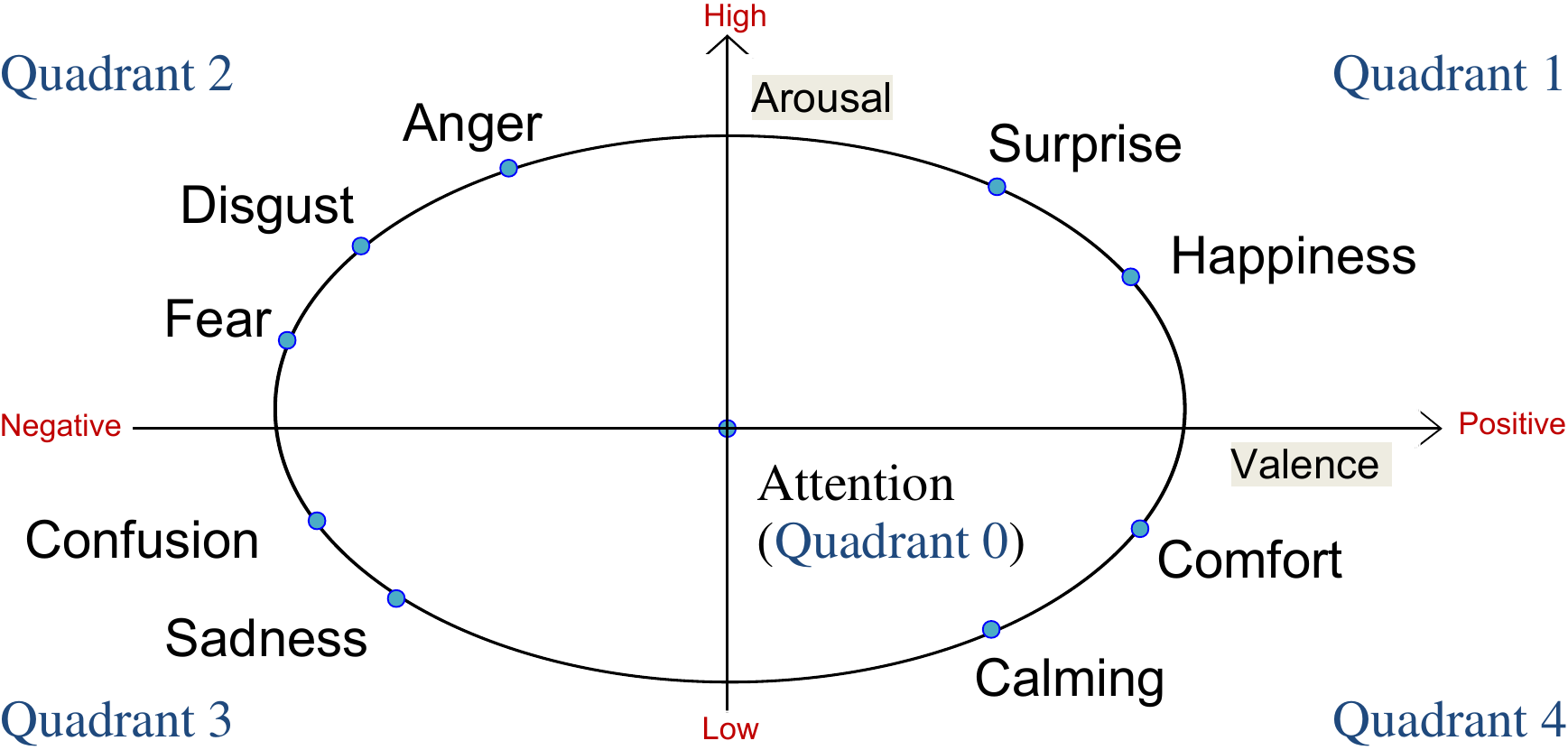}
    \caption{Circumplex Model \cite{russell1980circumplex} with 10 emotions in this paper.}
    \label{russel_model}
\end{figure}

\subsection{The proposed lightweight model: MTRCNN}\label{input_model}

\begin{figure}[H]
    \centering
    \setlength{\abovecaptionskip}{0.15cm}    
	\setlength{\belowcaptionskip}{-0cm} 
        \includegraphics [width=1\columnwidth]{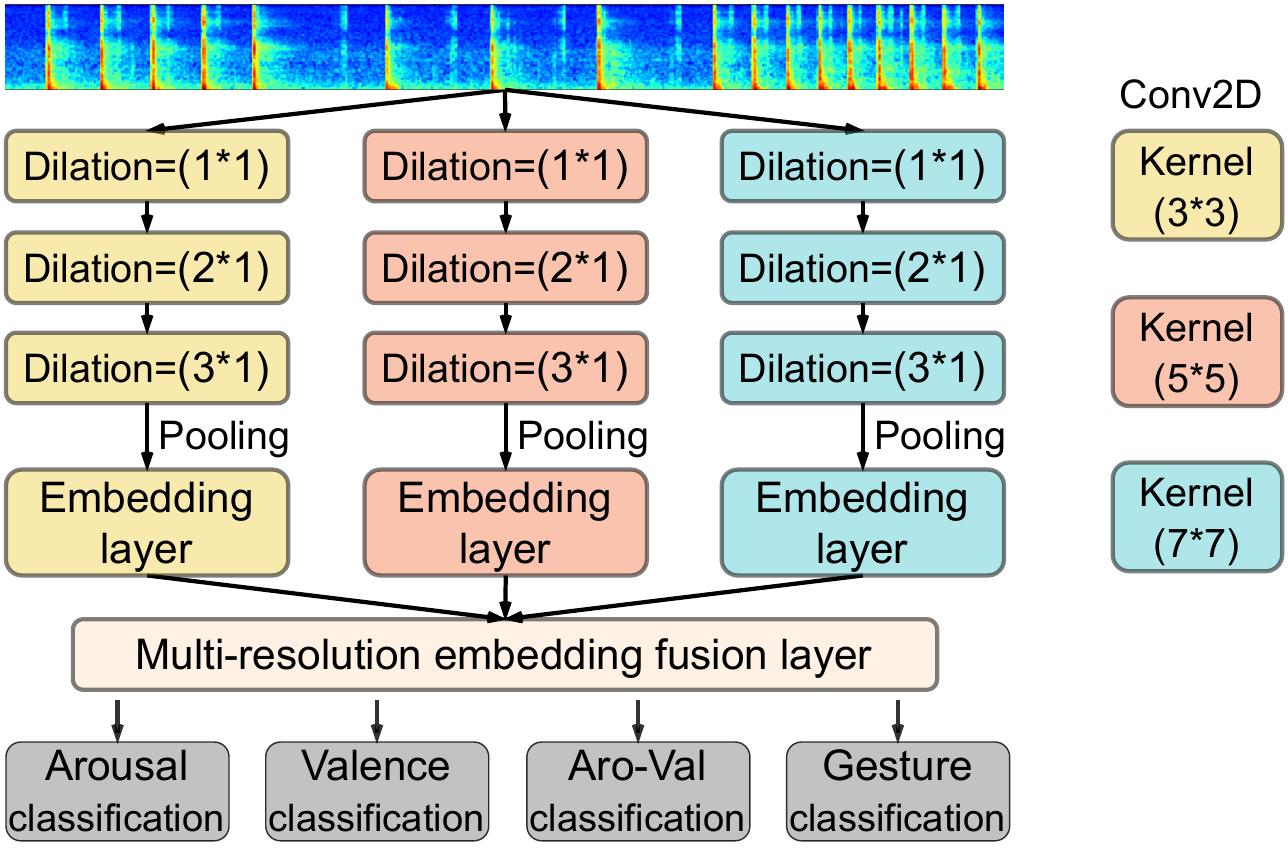}
    \caption{The proposed lightweight multi-temporal resolution convolutional neural network (MTRCNN).}
    \label{model}
\end{figure}

Sounds caused by various touch gestures and emotions may have different durations, e.g., the sound of an angry hit and the rust of a calm touch. 
Thus, we propose a multi-temporal resolution convolutional neural network (MTRCNN) in Fig.~\ref{model}. The convolution (Conv) part of MTRCNN consists of 3 branches with Conv kernel sizes of (3, 3), (5, 5), and (7, 7), respectively. Different kernel sizes extract representations with different resolutions. To obtain a larger Conv receptive field size (RFS) with fewer parameters, the dilated Conv \cite{HDC_dilated} is used. Moreover, to avoid the gridding artifacts \cite{gridding} of the dilated Conv, hybrid dilated Conv scheme \cite{HDC_dilated} is adopted, so the number of filters and the dilation rate of 3 Conv layers of each branch are [16, 32, 64] and [(1,1), (2,1), (3,1)]. The dilation rate only changes along the time axis.

\vspace{-0.1cm}

Taking the branch with the largest kernel (7, 7) as an example, according to the convolution  RFS calculation formula, 
\begin{equation}\label{formula_receptive_field}
\setlength{\abovedisplayskip}{3pt}
\setlength{\belowdisplayskip}{3pt}
R_i=(R_{i-1}-1) \times stride + k
\end{equation}
where $R_i$ is the $i$-th Conv layer's RFS relative to the input feature map, $R_0=1$, the Conv step size $stride$ defaults to 1, and $k$ is the Conv kernel size.  
If there is no pooling operation, according to Eq.~(\ref{formula_receptive_field}), the 1st Conv layer's RFS on the time axis is $R_1=7$. For dilated Conv, the formula for the RFS is
\begin{equation}
\setlength{\abovedisplayskip}{3pt}
\setlength{\belowdisplayskip}{3pt}
R_i=(R_{i-1}-1) \times stride + k + (k-1)(r-1)
\end{equation}
where $r$ is the dilation rate.
For the 2nd Conv layer with dilation rate (2, 1), on the time axis, $R_2$ is $20$. For the 3rd Conv layer, $R_3$ is $38$ on the time axis. That is, without pooling, MTRCNN requires that the length of the input be at least 38 frames. With a frame hop of 10\textit{ms}, the corresponding length of the input clip is at least 0.38\textit{s}. Identifying emotions or gestures within 0.38 seconds is challenging, even for humans. Moreover, if pooling is not used, the number of model parameters and the computation load will increase. After comprehensive trade-offs, we add pooling operations to these Conv layers, resulting in a minimum input audio length of 1.10\textit{s}.

After the three Conv layers, the representations are fed into the following 64D embedding layer to learn embeddings with different resolutions. Then, the 3 branches' embeddings are concatenated and fed into the 192-dimensional ($64*3$) multi-resolution embedding fusion layer to fuse information from different temporal resolutions. The fused embeddings are fed to the last classification layers. 
For the resulting MTRCNN model, the FLOPs is 0.708$G$, the number of parameters is 0.24$M$, the model size is 0.94$MB$, and it can process input audio clips of any length of at least 1.10$s$. These fit well with the design guidelines in Section (Sec.) \ref{guidelines}.

The tasks in subsequent experiments are all single-label multi-classification tasks, so the activation functions for the arousal (Aro), valence (Val), Aro-Val, and gesture classification layers are all Softmax, the loss function is cross entropy \cite{hou2023cooperative}. For more details, please see the \textbf{\textit{homepage}} (\textcolor{blue}{\url{https://github.com/Yuanbo2020/MTRCNN}}).

\section{Experiments and results}

\subsection{Dataset, experiments setup, and metrics}

We conducted data collection experiments to record touch sounds for gesture and touch sounds for emotion \cite{ren2024conveying}.  
As stated in Sec.~\ref{background}, participants first expressed gestures 10 emotions independently by touching Pepper's arm, and then expressed 6 touch gestures; the sounds generated by movements during touch are recorded as the dataset. Participants complete 3 rounds of interaction, each lasting 10\textit{s}. Finally, there are 84 ($28\times3$) 10\textit{s} audio clips for each gesture and emotion. For touch gesture classification, the number of samples in training, validation, and test sets is 366, 42, and 84, respectively. For emotion classification, the number of samples in training, validation, and test sets is 660, 80, and 100, respectively.

The Mel-filter with 64 banks is used as the acoustic feature, with a Hamming window of 32\textit{ms} and an overlap of 10\textit{ms} \cite{pann}. Dropout and normalization are used to prevent model overfitting \cite{dropout}. A batch size of 32 and Adam optimizer \cite{adam} with a learning rate of 1e-3 are used to minimize loss. Models are trained on a GPU card Tesla T4 for 100 epochs, and 10 times without a fixed seed to obtain the mean result of 10 runs. Accuracy (Acc) is used to evaluate the classification results. Dataset, code, and models are available on the \textbf{\textit{homepage}}.

\subsection{Results and analysis}
This part analyzes the performance of the proposed MTRCNN by the following research question (RQ).

\noindent
\textbf{RQ1}: Is it feasible to recognise touch gestures and distinguish emotions' arousal and valence based on audio alone?

Table~\ref{tab:RQ1} shows the results of MTRCNN for classifying 6 gestures, as well as arousal and valence of emotions. Arousal is usually classified as low, neutral, and high. Valence is classified as negative, neutral, and positive. The AVDM of arousal-valence joint classification has four quadrants and an origin, so the arousal-valence \cite{russell1980circumplex} joint classification has five categories: Q1, Q2, Q3, Q4, and Q0, as shown in Fig. \ref{russel_model}.

\begin{table}[H]\footnotesize
	\renewcommand\tabcolsep{1pt} 
	\centering
	\caption{Test set classification results of the model with 10s audio clip input; models are repeated 10 times.}
	\begin{tabular}{  
	p{0.9cm}<{\centering}|
	p{1.8cm}<{\centering}|
 p{1.8cm}<{\centering}|
	p{1.8cm}<{\centering}|
 p{1.8cm}<{\centering} }
	
		\toprule[1pt] 
		\specialrule{0em}{0.1pt}{0.1pt}  

    & 
  Arousal (Aro) & 
  Valence (Val) & 
  Aro-Val & 
  Gesture   \\
 
   
\hline 
	 Acc. &  69.97 $\pm$ 4.40 &  62.90 $\pm$ 4.78 & 53.93 $\pm$ 3.41 &  82.14 $\pm$ 3.46  \\

		\specialrule{0em}{0pt}{0em}
		\bottomrule[1pt]
	\end{tabular}
	\label{tab:RQ1}
\end{table}

\vspace{-0.2cm}
In Table~\ref{tab:RQ1}, MTRCNN performs better on touch gesture classification than emotion classification. This may be because gestures usually contain clear and consistent patterns, such as regular sounds when tapping and snapping sounds when patting, so the model can capture sounds produced by specific movements and rhythms to identify touch gestures effectively. However, due to the variety of ways in which different participants express emotions and different perceptions of the same type of emotions \cite{zadra2011emotion}, e.g., a calm emotion expressed by some may appear sad to others. This makes Aro-Val-based emotion classification, especially relying on sounds produced when touching, challenging.


In Table~\ref{tab:RQ1}, MTRCNN has a higher classification accuracy on the arousal dimension than the valence dimension. Arousal denotes the intensity of emotion, which is usually conveyed via direct physical cues, e.g., pressure, frequency, and speed, which allows MTRCNN to grab these cues in sounds to efficiently identify the class of arousal. 
Valence reflects emotion's positive or negative nature, which is more subtle and context-dependent \cite{de2012effect}, making it challenging to distinguish it based solely on touch actions and the sounds caused by it.

\noindent
\textbf{RQ2}: What is the shortest effective audio length required for touch-sound-based gesture and emotion recognition models?

Table \ref{tab:RQ1} shows the performance of MTRCNN trained with full 10\textit{s} audio clips. Here, we further explore the minimum audio length required for MTRCNN to effectively recognise gestures and emotion states. The input audio length can be regarded as a hyperparameter of the model. To avoid information leakage, Table \ref{tab:RQ2} shows the results of this hyperparameter on the validation set. 
As mentioned in Sec. \ref{input_model}, the proposed MTRCNN can handle audio clips with varying lengths with a minimum length of 1.10\textit{s}, so the input audio length range in Table \ref{tab:RQ2} is [1.10, 10].

Table \ref{tab:RQ2} shows that the accuracy of touch gesture classification increases with the input audio length and peaks at 6\textit{s}.

\begin{table*}[h]\footnotesize 
	\renewcommand\tabcolsep{1pt} 
	\centering
	\caption{Accuracy of Aro-Val joint classification and gesture classification with varying input lengths on the validation set}
	\begin{tabular}{  
	p{1.6cm}<{\centering}|
	p{1.55cm}<{\centering}|
    p{1.55cm}<{\centering}|
	p{1.55cm}<{\centering}|
    p{1.55cm}<{\centering}|
    p{1.55cm}<{\centering}|
    p{1.55cm}<{\centering}|
    p{1.55cm}<{\centering}|
	p{1.55cm}<{\centering}|
    p{1.55cm}<{\centering}|
    p{1.55cm}<{\centering} }
	
		\toprule[1pt] 
		\specialrule{0em}{0.1pt}{0.1pt}  

   Input length &  1.10 s  &  2.00 s  &  3.00 s   &  4.00 s  &  5.00 s  &  6.00 s  &  7.00 s  &  8.00 s  &  9.00 s  &  10.00 s  \\
 
   
\hline 
	 Aro-Val &  
  41.17$\pm$2.61 &  
  50.77$\pm$2.81 & 
  51.04$\pm$2.12 &  
  52.63$\pm$4.86 &
  53.29$\pm$4.26 & 
  
  54.46$\pm$3.71 & 
  \textbf{58.54}$\pm$3.74 &  
  56.58$\pm$5.62 & 
  53.83$\pm$3.46 &  
  51.79$\pm$3.84\\ 
  
  Gesture &  
  68.81$\pm$3.06 &  
  84.99$\pm$2.76 & 
  85.48$\pm$2.37 &  
  85.71$\pm$3.17 &  
  88.47$\pm$2.46 &
  
  \textbf{90.24}$\pm$3.80 &  
  86.19$\pm$3.33 &  
  83.81$\pm$3.69 & 
  83.57$\pm$6.39 &  
  79.31$\pm$6.55\\

		\specialrule{0em}{0pt}{0em}
		\bottomrule[1pt]
	\end{tabular}
	\label{tab:RQ2} 
\end{table*}

\noindent
The gesture classification results closest to the 6\textit{s} result are at 5\textit{s} and 7\textit{s}, respectively, so we conduct statistical analysis on these similar results. The Shapiro-Wilk test \cite{t_test} shows that the data follow a normal distribution. Then, the paired t-test \cite{paired_t} is used, and the statistics show that the gesture classification results based on 6\textit{s} clips are significantly better than those of 5s ($t=3.19, p<0.05$) and 7s ($t=3.22, p<0.05$). In addition, for the Aro-Val joint classification of emotions, the results closest to 7\textit{s} are the results of 6\textit{s} and 8\textit{s}, respectively. Paired t-test for Aro-Val joint classification shows that the results based on the 7\textit{s} clip are significantly better than those of 6s ($t=11.31, p<0.05$) and 8s ($t=2.58, p<0.05$).

The above analyses show that MTRCNN effectively recognises different touch gestures within 6\textit{s} and decodes emotions within 7\textit{s}. 
Hence, input lengths of 6\textit{s} and 7\textit{s} will be used as default settings of MTRCNN for touch gesture classification and Aro-Val classification of emotions, respectively.


\noindent
\textbf{RQ3:} What are the most challenging touch gestures and the emotions' dimensions to distinguish based on sounds?

\begin{figure}[t]
\vspace{-0.1cm}
    \centering
        \includegraphics [width=1\columnwidth]{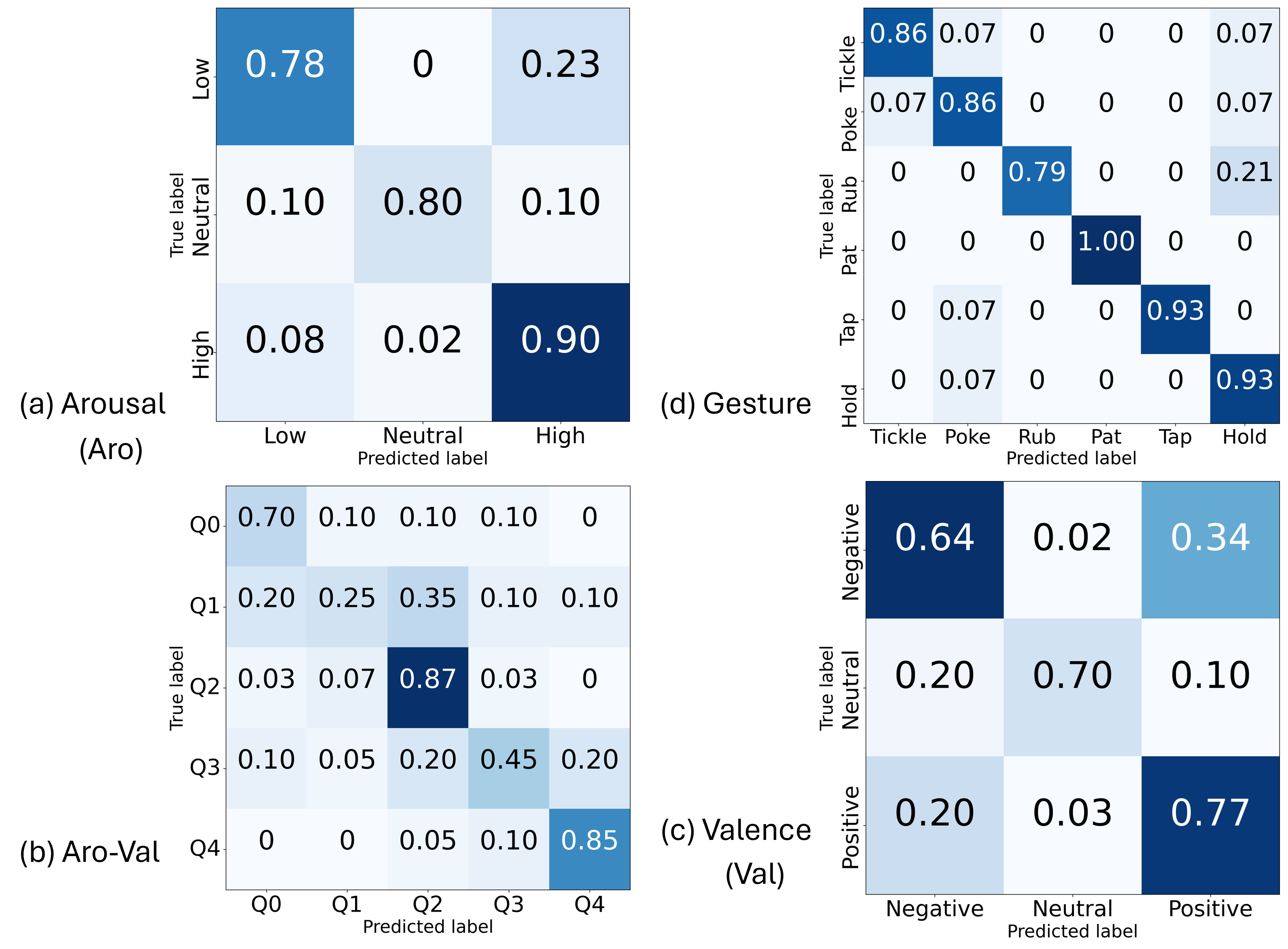}
    \caption{Normalized confusion matrix on the test set.}
    \label{confusion_matrix}
\end{figure}

In Fig.~\ref{confusion_matrix} (a), MTRCNN performs better in identifying high arousal than low and neutral states. This implies that high arousal associated with touch sounds is easier to distinguish. Fig. \ref{confusion_matrix} (c) implies that touch-based emotions in positive valence are more distinguishable than those in negative valence. In the Aro-Val space in Fig. \ref{confusion_matrix} (b), MTRCNN can better distinguish emotions in Q2 (high arousal, negative valence) and Q4 (low arousal, positive valence) than those in Q1 and Q3. This is interesting because it suggests that these combinations of arousal and valence may be more consistently expressed by specific touch gestures conveyed by participants. 
The high and low arousal in Fig.~\ref{confusion_matrix} (a), as well as positive and negative valence in Fig.~\ref{confusion_matrix} (c), are rarely misclassified as neutral, implying that non-neutral emotions are less likely to be confused with neutral emotion due to their different tactile cues.

For gesture recognition, as shown in Fig.~\ref{confusion_matrix} (d), MTRCNN accurately identifies gestures such as \textit{pat}, \textit{tap}, and \textit{hold}. These gestures usually have unique tactile profiles that are easy to identify. Like, \textit{pat} may involve an easily recognizable repetitive rhythmic pattern \cite{hogan2007rhythmic}, while \textit{tap} and \textit{hold} are simple, discrete movements with clear tactile features. In addition, \textit{rub} is only misclassified as \textit{hold}. This may be due to the subtle differences and overlapping tactile sensations between the two gestures, making their sounds similar, especially when the strength or speed of \textit{rub} is not obvious.

\noindent
\textbf{RQ4}:  How does the proposed model perform compared to other typical sound-related models?

Table \ref{tab:other_models} compares models' performance on the same server (CPU: AMD EPYC 7352, GPU: Tesla T4 16GB). CNN-Transformer consists of 3 convolutional layers with (3 × 3) kernels, a Transformer encoder, and the final classification layers. YAMNet and MobileNetV2 \cite{sandler2018mobilenetv2} are classic CNN-based networks. PANNs \cite{pann} have shown excellent performance on AudioSet \cite{audioset} and audio pattern-related tasks. 
Therefore, \#4 and \#5 explore the performance of PANNs with and without pretrained weights on large-scale AudioSet \cite{audioset}, respectively. The \#5 with pretrained weights (PreW) significantly outperforms \#4, indicating that the knowledge from the 5800-hour AudioSet effectively improves the model performance. However, \#6, which has the smallest number of parameters and model size, achieves the best results in gesture classification, and is better than PANNs with PreW in \#5. For Aro-Val classification, although \#5, which uses PreW from AudioSet, is slightly better than the proposed \#6 on the mean values, there is no statistically significant difference ($t=-0.70, p>0.05$).

\begin{table}[H] \footnotesize 
	\renewcommand\tabcolsep{1pt} 
	\centering
	\caption{Comparison of different models on the test set.}
	\begin{tabular}{
	p{0.2cm}<{\centering}| 
 p{1.7cm}<{\centering}| 
 p{0.75cm}<{\centering} |
	p{0.7cm}<{\centering}|
 p{0.82cm}<{\centering}|
 p{1.05cm}<{\centering}|
 p{1.4cm}<{\centering}|
 p{1.4cm}<{\centering}
	} 
	   \toprule[1pt] 
    \specialrule{0em}{0.1pt}{0.1pt} 

  \multirow{2}{*}{\makecell[c]{\#}} & \multirow{2}{*}{\makecell[c]{Model}} &  Param. &  

Size & FLOPs & Inference &  \multicolumn{2}{c}{Accuracy} \\

\cline{7-8} 

  & & (M) & (MB) & (G) & time (s) & \multirow{1}{*}{\makecell[c]{Aro-Val}} & \multirow{1}{*}{\makecell[c]{Gesture}} \\

  \hline 
   
1 & CNN-Trans.  & 1.58 & 6.02  &  0.266 &  0.006  &  28.87$\pm$4.47 & 60.60$\pm$6.51  \\

2 & YAMNet  & 3.21 & 12.30  &  0.728 & 0.008 &  29.03$\pm$2.77  & 61.19$\pm$5.78 \\

  3 & MobileNetV2  & 2.23 &  8.74  &  0.351 &  0.007 &  45.63$\pm$6.22 &  71.90$\pm$5.56   \\

 4 & PANNs & 79.68 &  304.1  & 11.96 &  0.012 &  49.20$\pm$3.02 &  76.43$\pm$4.48 \\

 5 & PANNs PreW & 79.68 &  304.1 &  11.96 &  0.012  & 55.83$\pm$4.84 & 83.33$\pm$3.12  \\

\hline
 6 & MTRCNN & \textbf{0.24} &  \textbf{0.94} & 0.708 & 0.007 & 54.73$\pm$3.29 & \textbf{84.17}$\pm$3.89\\
	 
	\specialrule{0em}{0pt}{0em}
		\bottomrule[1pt]
	\end{tabular}
	\label{tab:other_models}
\end{table}

\section{CONCLUSION}
\label{sec:CONCLUSION}

This paper explores the feasibility of identifying touch gestures and the emotions leading to them based on sounds produced by movements during touch, which fills the gap in HRI that lacks touch-related sounds to decode touch gestures and emotions. The proposed sound-based touch gesture and emotion recognition model can effectively recognize the arousal and valence states of different emotions, as well as various tactile gestures, when the input audio length varies from at least about 2 seconds to the optimal 6 to 7 seconds. Moreover, its lightweight, low-parameter, and low-latency processing characteristics make it ideal for real-time applications on robots such as Pepper. Future work will package the proposed model into an application and deploy it on Pepper.

\vfill\pagebreak

\label{sec:refs}

\bibliographystyle{IEEEbib}
\bibliography{main}

\end{document}